# Coherent modulation up to 100 GBd 16QAM using silicon-organic hybrid (SOH) devices

S. WOLF,[1] H. ZWICKEL,[1] C. KIENINGER,[1,2] M. LAUERMANN,[1,3] W. HARTMANN,[1,2,4] Y. KUTUVANTAVIDA,[1,2] W. FREUDE,[1] S. RANDEL,[1] AND C. KOOS[1,2,*]

[1]*Karlsruhe Institute of Technology (KIT), Institute of Photonics and Quantum Electronics (IPQ), Karlsruhe, Germany*
[2]*Karlsruhe Institute of Technology (KIT), Institute of Microstructure Technology (IMT), Karlsruhe, Germany*
[3]*Now with: Infinera Corporation, Sunnyvale, CA, United States*
[4]*Now with: Physikalisches Institut, University of Münster, Muenster, Germany*
*\*christian.koos@kit.edu*

**Abstract:** We demonstrate the generation of higher-order modulation formats using silicon-based inphase/quadrature (IQ) modulators at symbol rates of up to 100 GBd. Our devices exploit the advantages of silicon-organic hybrid (SOH) integration, which combines silicon-on-insulator waveguides with highly efficient organic electro-optic (EO) cladding materials to enable small drive voltages and sub-millimeter device lengths. In our experiments, we use an SOH IQ modulator with a π-voltage of 1.6 V to generate 100 GBd 16QAM signals. This is the first time that the 100 GBd mark is reached with an IQ modulator realized on a semiconductor substrate, leading to a single-polarization line rate of 400 Gbit/s. The peak-to-peak drive voltages amount to 1.5 $V_{pp}$, corresponding to an electrical energy dissipation in the modulator of only 25 fJ/bit.

## References and Links

## 1. Introduction

Advanced modulation formats such as quadrature phase-shift keying (QPSK) and 16-state quadrature amplitude modulation (16QAM) are widely deployed in metro and long-haul networks. This has significantly contributed to increase the data rates that can be transmitted over a single wavelength-division multiplexing (WDM) channel. However, it is additionally necessary to increase the underlying symbol rates, while the underlying devices must be amenable to compact and cost-efficient integration to maintain economical and technical scalability of high-performance WDM systems to large channel counts [1].

In laboratory experiments, net data rates up to 1 Tbit/s on a single optical carrier were demonstrated using QAM modulation at symbol rates up to 100 GBd [2,3]. However, these experiments relied on conventional lithium-niobate modulators, which lack the potential for dense photonic integration and which were combined with an optical equalizer to compensate the low-pass characteristics of the device. As a promising alternative, modulators based on semiconductors like indium-phosphide (InP) or silicon (Si) have been explored, offering comparable bandwidths, but much smaller footprint. Using high-bandwidth InP modulators, the generation of single-carrier signals at line rates (net data rates) of up to 352 Gbit/s (286 Gbit/s) has been demonstrated [4,5]. However, while InP offers the potential to monolithically integrate modulators and lasers, it requires expensive processing and is not amenable to co-integration of electronic circuits.

In contrast to that, silicon photonics [6–10] leverages advanced large-scale cost-efficient complementary metal-oxide-semiconductor (CMOS) fabrication processes [9,11], which are designed for high integration density and high yield, and which offer a path towards co-integration of photonic circuits with electronics [12]. However, up to now, conventional all-silicon photonic (SiP) IQ modulators still lag behind their InP counterparts, both in terms of transmission speed and efficiency. The highest reported line rate for SiP modulators amounts to 227 Gbit/s per polarization, exploiting electrical orthogonal frequency-division multiplexing (eOFDM) [13]. In this experiment, a forward-error correction (FEC) algorithm with a code rate of 0.68 (47 % overhead) was used to achieve error-free transmission after 480 km, thus leading to a net data rate of 154.2 Gbit/s. In another experiment, a net data rate of 150 Gbit/s was generated using a 30 GBd polarization-division multiplexing (PDM) 64QAM signal [7]. Hence, up to now, using SiP IQ modulators, the 400 Gbit/s mark, corresponding to a single-polarization net data rate of 200 Gbit/s, has not yet been reached. Moreover, all these devices rely on carrier depletion in reversely biased *p-n* junctions [14], which leads to a comparatively low modulation efficiency. In general, the modulation efficiency of EO modulators is quantified by the π-voltage-length product $U_\pi L$, which typically amounts to a relatively large value of the order of 10 Vmm for depletion-type SiP MZM [15–17]. As a consequence, broadband drive amplifiers providing voltage swings between 2.5 $V_{pp}$ and 5 $V_{pp}$ are required to operate these devices [18,19].

Silicon-organic hybrid (SOH) modulators [20,21] overcome the limited efficiency of all-silicon devices by combining conventional silicon-on-insulator (SOI) waveguides with highly-efficient electro-optic organic cladding materials [22,23]. The underlying SiP waveguide structures can be fabricated using optical lithography, and the organic materials are deposited in a back-end-of-line process. With voltage-length products down to 0.5 Vmm [24,25], the efficiency of SOH Mach-Zehnder modulators (MZM) surpasses the efficiency of all-silicon devices by more than one order of magnitude. Using SOH MZM, on-off-keying (OOK) signals have been generated with drive voltages of only 80 $mV_{pp}$ [25], and the devices are also well suited for higher-order modulation formats [26–28]. The generation of 112 Gbit/s 16QAM signals has been demonstrated at a voltage swing of only 600 $mV_{pp}$ without any drive amplifiers [26]. SOH IQ modulators can also be operated directly from the binary outputs of a field-programmable gate array (FPGA), where line rates of up to 52 Gbit/s were demonstrated [27]. Besides low drive voltages, SOH EO modulators allow high-speed signaling for both coherent and direct-detection transmission [28–30].

In this paper, we demonstrate that the SOH concept allows to further push the limits of silicon-based IQ modulators by enabling 16QAM signaling at symbol rates of up to 100 GBd. In a first set of experiments, we use CMOS digital-to-analog converters (DAC) to drive an SOH IQ at 16QAM symbol rates of 63 GBd, thus leading to a single-polarization line rate of 252 Gbit/s [31]. The bit error ratio (BER) amounts to $4.1 \times 10^{-3}$ such that forward error correction (FEC) scheme with a 7 % overhead can be applied, thus leading to an error-free reception with a net data rate of 232 Gbit/s. In a second set of experiments, we use BiCMOS DAC and employ a bandwidth-increasing gate field [32–34] to demonstrate 16QAM signaling at line rates of up to 100 GBd, corresponding to single-polarization line rates of up to 400 Gbit/s. For the 100 GBd 16QAM demonstration, we obtain BER values that are below the limits of state-of-the-art third-generation FEC algorithms. Considering a code rate of 0.83 (20 % overhead), the line rate of 400 Gbit/s corresponds to a net data rate of more than 330 Gbit/s [35]. This represents the highest line rate and the highest net data rate so far demonstrated for an integrated EO modulator fabricated on a semiconductor substrate. Exploiting the second polarization would enable single-carrier line rates of up to 800 Gbit/s. The 63 GBd and 100 GBd signals were generated with small peak-to-peak drive voltages of 1.0 $V_{pp}$ and 1.5 $V_{pp}$, respectively – the lowest value so far demonstrated for semiconductor-based IQ modulators at these speeds. The electrical energy consumption of the modulator amounts to only 25 fJ/bit for the 400 Gbit/s signal – more than an order of magnitude below that of conventional all-silicon modulators at considerably lower speeds [18,19]. These proof-of-principle experiments underline the outstanding properties of SOH modulators and show a path towards coherent silicon-based transmitters operating at line rates of 400 Gbit/s and above without the need for power-hungry drive amplifiers.

## 2. Silicon-Organic Hybrid (SOH) IQ Modulator

Silicon-organic hybrid (SOH) modulators rely on the interaction of the optical wave with an EO organic cladding material in a silicon-on-insulator (SOI) slot waveguide [20]. The concept of an SOH IQ modulator is depicted in Fig. 1 (a). The IQ modulator consists of two nested Mach-Zehnder (MZ) interferometers formed by SOI waveguides and multi-mode interference (MMI) couplers, which split or combine the fields at their inputs. A cross-section A - A' of a single SOH MZM is depicted in Fig. 1 (b). The phase shifter sections are realized as SOI slot waveguides which are formed by two 240-nm wide silicon (Si) rails that are separated by the 120-nm wide slots. The slots are filled with an organic electro-optic cladding material. For the fundamental quasi-TE mode (dominant transverse electric field component parallel to the substrate), the high refractive-index contrast between waveguide and EO cladding leads to a strong confinement inside the slot [36], as visualized in Fig. 1(c). Utilizing the high linear electro-optic coefficient of the organic cladding material, the refractive index is modulated by an electric field between the two silicon rails. The modulating electric field is generated by applying a drive voltage, $U_{drive}(t)$ to coplanar ground-signal-ground (GSG) travelling-wave electrodes, which are connected to the silicon rails via 70-nm high slightly *n*-doped conductive slabs, see Fig. 1(b). This design ensures that the modulating electric field $E_{x,RF}(t)$ is also confined in the narrow slot, Fig. 1(d), thus leading to a strong overlap of the optical and the modulating electric fields and resulting in a high modulation efficiency with $U_{\pi}L$ products down to 0.5 Vmm [25]. The silicon waveguides and the metallization layer are structured using optical lithography in a standard 248 nm deep-UV (DUV) process at A*Star IME, Singapore, and are connected through aluminum (Al) vias. For the deposition of the EO material, the slots are accessed by locally opening the oxide in the respective regions. In order to achieve EO activity in the cladding, the material is poled in a one-time procedure after deposition. To this end, the device is first heated close to its glass-transition. A poling voltage $U_{pol}$ is then applied across the floating ground electrodes, see Fig. 1(b), such that half of the voltage drops across each of the slots. This leads to an alignment of the dipolar chromophores in the slot in the direction of the poling field, indicated by the green arrows in Fig. 1 (b). Finally, the device is cooled down to ambient temperatures while holding the poling voltage. The chromophore alignment is thus "frozen" and will remain, even after removing the poling voltage. Applying the modulating voltage $U_{drive}(t)$ to the GSG electrodes then leads to electric fields, indicated by red arrows, Fig. 1 (b). These fields are oriented in the same (opposite) direction as the chromophores in the right (left) slot. Consequently, the phase shifts in both arms have equal magnitude but opposite sign, and the SOH MZM operates in push-pull mode when applying a single-ended drive signal, allowing nearly chirp-free modulation [30].

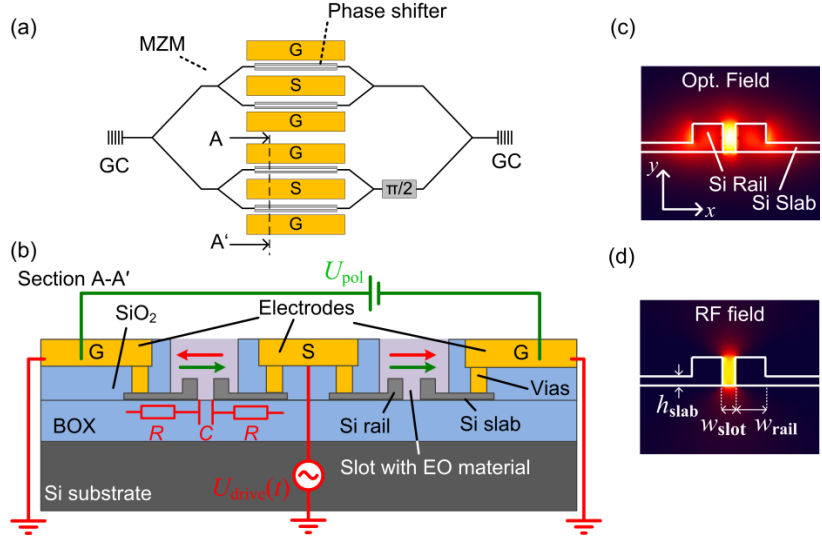

Fig. 1. Concept of the silicon-organic hybrid (SOH) modulator. **(a)** Schematic of an SOH inphase-quadrature (IQ) modulator consisting of two nested Mach-Zehnder modulators (MZM). The thin black lines represent standard silicon-on-insulator (SOI) strip waveguides. The phase modulator sections are based on slot waveguides (rail width $w_{rail}$ = 240 nm, slot width $w_{slot}$ = 120 nm) and represented by rectangles in light grey. Coplanar ground-signal-ground (GSG) radio-frequency transmission lines carry the modulation signals. **(b)** Schematic of the SOH MZM cross section along the line AA' illustrated in (a). The transmission line is electrically connected to the slot by aluminum (AL) vias and thin $n$-doped silicon slabs (thickness $h_{slab}$ = 70 nm, width 8 µm). The chip is overclad with SiO$_2$, which is locally removed in the slot areas. The slots are the covered with an organic electro-optic (EO) material, which is deposited on the chip such that it homogeneously fills the slots. The chromophores are aligned (green arrows) at an elevated temperature using a poling voltage $U_{pol}$ applied to the floating ground (G) electrodes. After cooling to ambient temperature, the orientation of the chromophores is frozen, and the poling voltage can be removed. For operation of the device, the modulating signal is applied to the GSG line. This RF field (red arrows) is oriented in opposite direction to the chromophore alignment in one arm, and in the same direction in the other arm of the MZM such that two phase modulators of the MZM are operated in push-pull mode. **(c)** Fundamental quasi-TE mode of the slot waveguide. The color-coded graph shows the magnitude of the dominant optical electric field (dominant $x$-component), which is strongly confined to the slot. **(d)** Magnitude of the electrical RF modulation field (dominant $x$-component), which is also strongly confined to the slot. The good overlap of optical and RF field results in efficient EO modulation.

The modulation bandwidth of SOH modulators is limited by an RC characteristic resulting from the resistivity of the silicon slabs (represented by a lumped resistor $R$) and the capacitance of the slot (represented by a lumped capacitance $C$). Previous high-speed experiments with SOH modulators [28,29,37] relied on the use of a gate voltage applied between the bulk silicon and the device layer. The gate voltage leads to a highly conductive electron accumulation below the Si slab surface near the SiO$_2$-layer, which decreases the slab resistivity and consequently increases the modulation bandwidth [32,34]. In future designs, an improved lateral doping profile of the 8-µm wide silicon slabs could eliminate the need for a gate voltage without compromising the optical attenuation.

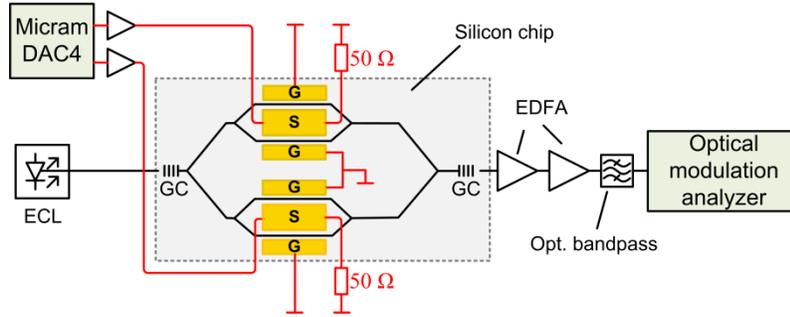

Fig. 2. Experimental setup. An arbitrary-waveform generator (AWG) is used for driving the modulator via radio-frequency (RF) amplifiers. The electrical drive signals are fed to the ground-signal-ground (GSG) transmission lines of the SOH IQ modulator. The transmission lines are terminated by external 50 Ω resistors. The optical carrier is provided by an external cavity laser (ECL). Grating couplers (GC) are used to couple light to and off the chip. The insertion loss of the modulator is compensated by an erbium-doped fiber amplifier (EDFA), and an optical bandbass filter is used to suppress out-of-band amplified spontaneous emission (ASE) noise. The signal is received, recorded and evaluated using an optical modulation analyzer (OMA). We perform two sets of experiments, which we refer to as Experiment 1 and Experiment 2 in the following. Both experiments use the same setup scheme, but differ in the specifications of the actual equipment, in particular in the bandwidth of the AWG and the OMA, see Appendix A for details.

In the following sections, we describe two sets of high-speed modulation experiments, obtained with and without a gate voltage. First without a gate field, we show modulation and detection at net rates of up to 235 Gbit/s (252 Gbit/s line rate) in a single polarization at an energy consumption of down to 22 fJ/bit. In a second set of experiments, we use a gate voltage to push the modulator limits, breaking the 100 GBd mark. In this experiment, single-polarization optical signals with a net rate of 333 Gbit/s (400 Gbit/s including FEC overhead) are generated at a slightly higher energy consumption of 25 fJ/bit.

### 3. Setup for optical data generation

The basic experimental setup is depicted in Fig. 2. The electrical drive signals are derived from an arbitrary-waveform generator (AWG) and rely on random bit sequences. Radio-frequency (RF) amplifiers are used to increase drive voltage levels. The GSG transmission lines of the SOH IQ modulator are fed using a microwave probe and terminated by external 50 Ω resistors that are connected via a second probe. The optical carrier is provided by an external cavity laser (ECL) and coupled to and off the chip via grating couplers (GC). An erbium-doped fiber amplifier (EDFA) after the modulator compensates the insertion loss and acts as a receiver pre-amplifier, and an optical bandpass filter is used to remove out-of-band amplified spontaneous emission (ASE) noise. The signal is received by an optical modulation analyzer (OMA) and evaluated off-line. In the experiments, the length of the signal recording is limited by the available hardware. The bit error ratio (BER) can be measured directly in case enough errors are found within the signal recordings. Otherwise, we use the error vector magnitude (EVM) [38] as quality metric. The EVM describes the deviation of a received constellation point from its ideal position in the complex plane and can be used to estimate the BER under the assumption that the signal is distorted by additive white Gaussian noise only [38,39].

Using the scheme depicted in Fig. 2, we performed two sets of experiments, which are referred to as Experiment 1 and Experiment 2 in the following. Both experiments use the same setup scheme, but differ in the specifications of the actual equipment, in particular in the analog bandwidth of the electronics at the transmitter and receiver, see Appendix A for a detailed list and the technical specifications of the various equipment items. The two investigated SOH modulators have virtually the same specifications: In the first (second) experiment, the phase shifter length of the IQ modulator amounts to 0.5 mm (0.6 mm), and $U_\pi L$-products of 1.1 Vmm (1.0 Vmm) were achieved by using SEO100 (SEO250) as EO material [40–43]. Note that these $U_\pi L$-products are slightly larger than the ones reported in previous work [24,25] because the EO materials were chosen for good temperature stability rather than for highest EO activity. For SEO100, EO coefficients of $r_{33} = 166$ pm/V have been achieved in bulk material, and the glass transition temperature amounts to 140 °C. [42]. Using this material, we have previously demonstrated device operation at temperatures of 80 °C [37].

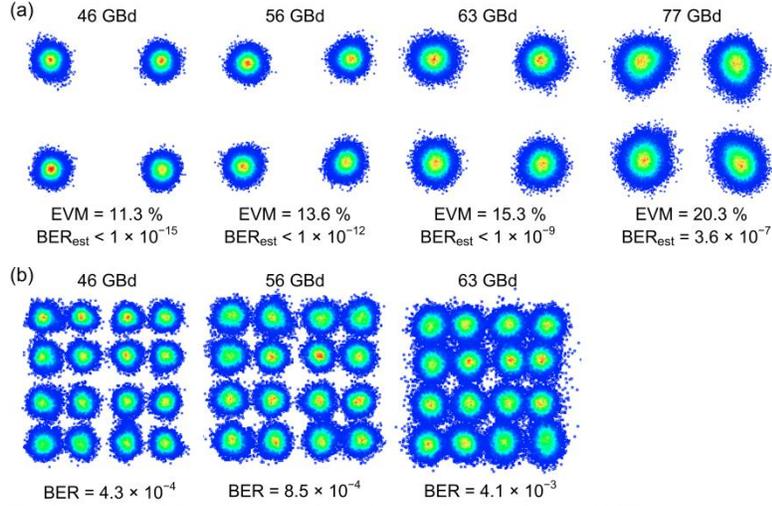

Fig. 3. Results of Experiment 1. Constellation diagrams for (a) QPSK and (b) 16QAM signals with symbol rates of up to 63 GBd for 16QAM signaling and up to 77 GBd for QPSK. In the QPSK experiment, we measure the error vector magnitude (EVM) and estimate a corresponding bit error ratio ($BER_{est}$) [38]. With estimated BER values below $10^{-9}$, the QPSK signal can be considered error free for symbol rates up to 63 GBd. For the 77 GBd QPSK signal with a line rate of 154 Gbit/s on a single polarization, we measure an EVM of 20.3 % corresponding to an estimated BER of $3.6 \times 10^{-7}$ – still well below the threshold of hard-decision FEC with 7 % overhead. For the 16QAM signals, the BER is directly measured from the signal recordings. The BER values range from $4.3 \times 10^{-4}$ for the 46 GBd signal to $4.1 \times 10^{-3}$ for the 63 GBd signal representing a line rate of 252 Gbit/s. All 16QAM BER values are hence well within the limits of hard-decision FEC with 7 % overhead.

In the first experiment, the insertion loss of the SOH IQ modulator chip amounts to approximately 17.5 dB, dominated by the fiber-chip coupling losses of approximately 8 dB (4 dB per GC-interface). The device used in the second experiment had a higher insertion loss due to a specific chip design which contained additional power splitters in the feeding waveguide sections. We expect that the total fiber-to-fiber insertion losses (on-chip losses) of the IQ modulator can be reduced to significantly less than 10 dB (5 dB) by improved fabrication processes that reduce sidewall roughness, by asymmetric slot waveguide geometries [44], and by optimized doping profiles for the phase shifter sections [28].

## 4. Coherent signaling generation up to 400 Gbit/s

In Experiment 1, we use an AWG featuring an analog bandwidth of 32 GHz and a sampling rate of 92 GSa/s (Keysight M8196A) along with an OMA featuring an analog bandwidth of 63 GHz and a sampling rate of 160 GSa/s (Keysight DSOZ634A). We first generate QPSK signals with data rates up to 154 Gbit/s, corresponding to symbol rates of up to 77 GBd – the corresponding constellation diagrams are shown in Fig. 3 (a). The measured EVM values of 11.3 % and 13.6 % for symbol rates of 46 GBd and 56 GBd correspond to estimated BER values well below $10^{-15}$ and $10^{-12}$, respectively. The 63 GBd signal can still be considered error-free as we estimate a BER $< 10^{-9}$ from the EVM of 15.3%. For symbol rates of 73 GBd and 77 GBd, the measured EVM of 19.7% and 20.3% correspond to BER values of $2 \times 10^{-7}$ and $3.6 \times 10^{-7}$, respectively. A direct measurement of the BER was not possible for the QPSK signals as no errors were measured in our recording of $6.5 \times 10^6$ bits. Still, these values are clearly below the $4.5 \times 10^{-3}$ limit for hard-decision forward error correction (FEC) codings [45] with 7 % overhead. The received signals were evaluated using Keysight's vector signal analysis (VSA) software [46] using a series of digital processing stages such low-pass filtering, polarization demultiplexing, chromatic dispersion compensation, frequency offset estimation, carrier phase estimation and adaptive equalization.

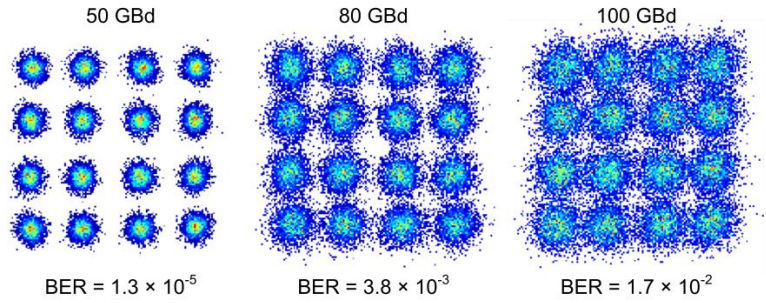

Fig. 4. Results of Experiment 2: 16QAM constellation diagrams for symbol rates of 50 GBd (left), 80 GBd (center) and 100 GBd (right). The BER up to 80 GBd (320 Gbit/s) remains below the hard-decision FEC limit with 7 % overhead, while the 100 GBd (400 Gbit/s) signal is below the soft-decision FEC threshold with 20 % overhead. The net data rate is larger than 330 Gbit/s on a single polarization.

The 16QAM constellation diagrams are depicted in Fig. 3 (b) - here we measure BER values directly. The lowest BER value is $8 \times 10^{-6}$ for a 23 GBd 16QAM signal corresponding to a line rate of 92 Gbit/s (not depicted). For the 46 GBd and 56 GBd (184 Gbit/s and 224 Gbit/s) signals, the measured BER is $4.3 \times 10^{-4}$ and $8.5 \times 10^{-4}$, respectively. For the highest symbol rate of 63 GBd, corresponding to a line rate of 252 Gbit/s for a 16QAM modulation, the BER is $4.1 \times 10^{-3}$. All of the measured BER values are within the $4.5 \times 10^{-3}$ limit for hard-decision FEC coding with 7 % overhead. Considering the FEC overhead, the single-polarization line rates of 154 Gbit/s for the 77 GBd QPSK signal and of 252 Gbit/s for the 63 GBd 16QAM signal correspond to net data rates of 143 Gbit/s and 234 Gbit/, respectively.

In Experiment 2, we expand the analog bandwidths of both the AWG and the OMA for increasing the symbol rate of the 16QAM signaling. As an AWG, we use a Micram DAC4 featuring an analog bandwidth of more than 40 GHz and a sampling rate of up to 100 GSa/s [47]. The OMA comprises a coherent receiver (Tektronix OM4245) featuring an analog bandwidth of 45 GHz and 70 GHz real-time oscilloscopes with sampling rates of 200 GSa/s (Tektronix DPO77002SX). We further apply a gate field [32–34] of 0.1 V/nm to increase the modulator bandwidth. The larger transmitter bandwidth allows for symbol rates up to 100 GBd at the cost of a higher noise level. This requires advanced FEC schemes with BER thresholds of $2 \times 10^{-2}$ or even higher, which may use soft-decision or staircase architectures [48–50]. We further use dedicated digital signal processing (DSP) at the receiver, including digital timing recovery along with a 33-tap fractionally spaced (two-fold oversampled) adaptive feed-forward equalizer that is adapted by a least-mean-square stochastic gradient algorithm [51]. To this end, we use training symbols to initialize the coefficients, before we switch to a decision-directed operation. We compute the BER from a sample of more than 500 000 bit.

The results for the 16QAM generation are depicted in Fig 4. At 50 GBd (200 Gbit/s) and 80 GBd (320 Gbit/s), we measure a BER of $1.3 \times 10^{-5}$ and $3.8 \times 10^{-3}$, respectively. The measurement shows an improvement with respect to the previous results, enabled by the bandwidth improvements in the setup and by the optimized DSP. Both BER values are within the threshold of a hard-decision FEC with 7 % overhead, leading to net data rates of 186 Gbit/s and 299 Gbit/s. At 100 GBd, corresponding to a line rate of 400 Gbit/s on a single polarization, the measured BER of $1.7 \times 10^{-2}$ is within the limits of today's soft-decision FEC codes. Considering a FEC overhead of 20%, this results in a net data rate of more than 330 Gbit/s. Note that, in contrast to Experiment 1, the 100 GBd signals are generated without oversampling such that pulse shaping techniques cannot be used. For a fair comparison of the results, the 50 GBd signal (2-fold oversampling) and the 80 GBd signal (no oversampling) are also generated without pulse shaping. To the best of our knowledge, our experiments represent the first demonstration of 100 GBd signaling using an IQ modulator realized on a semiconductor substrate.

## 5. Energy considerations

Power dissipation is an important aspect in today's optical communication networks. Especially the electronic drivers for EO modulators contribute significantly to the total power consumption of a transceiver unit. The specifications for the drivers are dictated by the choice of the EO modulator and the required drive voltages. As a figure of merit which is related to the transceiver's energy consumption, we use the power dissipation in the modulator.

For an estimate of the IQ modulator power dissipation, we assume that the transmission line impedances of the two MZM are perfectly matched to the terminating resistances $R = 50\ \Omega$ and to the 50 $\Omega$ internal resistances of the driver. We further assume that all symbols of the 16QAM-constellation occur with equal probability. The energy per bit then depends on the line rate $r$ and the peak-to-peak drive voltage $U_{\text{drive}}$ as measured from the electrical eye opening [26],

$$W_{\text{bit}}^{(16\text{QAM})} = 2\left[\frac{1}{2}\left(\frac{U_{\text{drive}}}{2}\right)^2 + \frac{1}{2}\left(\frac{1}{3}\frac{U_{\text{drive}}}{2}\right)^2\right]\frac{1}{R}\times\frac{1}{r}.$$

The 100 GBd 16QAM signal with a line rate of 400 Gbit/s was generated with a peak-to-peak drive voltage of $U_{\text{drive}} = 1.5\ V_{\text{pp}}$, the energy per bit is 25 fJ/bit. For the 252 Gbit/s experiment with a lower voltage swing of $U_{\text{drive}} = 1\ V_{\text{pp}}$, we find a slightly reduced energy consumption of 22 fJ/bit.

## 6. Summary

We demonstrate high-speed coherent signaling using silicon-organic hybrid (SOH) IQ modulators. The SOH approach expands the capabilities of the silicon photonic integration platform by combination with highly efficient organic electro-optic (EO) materials, thereby enabling highly efficient high-speed optical phase modulation without amplitude-phase coupling. In our experiments, we use SOH modulators with π-voltage-length products of approximately $U_\pi L = 1$ Vmm to generate data streams with line rates of up to 400 Gbit/s. Up to line rates of 320 Gbit/s, the measured BER remains within the limits of hard-decision forward-error correction, resulting in a net data rate of 299 Gbit/s. At 400 Gbit/s, the BER is still below the threshold for soft-decision FEC and allows a net transmission of more than 330 Gbit/s. Our experiments demonstrate the highest symbol rates and the highest data rates reported so far for IQ modulators that are realized on a semiconductor platform, thereby emphasizing the unique advantages of the SOH integration for high-speed energy-efficient coherent communications.

## Appendix A

This section gives a detailed overview on the equipment used in the two experiments, which are referred to as Experiment 1 and Experiment 2 in the main text of the manuscript.

**Experiment 1:** The signal is generated by an arbitrary-waveform generator (AWG, Keysight M8196A), featuring an analog bandwidth of 32 GHz and a sampling rate of up to 92 GSa/s. The data pattern is derived from a pseudo-random binary sequence (PRBS) with a length of $2^9 - 1$. As a pulse shape we choose a raised-cosine (RC) with roll-off factor $\beta = 0.35$. We generate QPSK signals at symbol rates of up to 77 GBd and 16QAM signals up to 63 GBd. For signal generation up to 63 GBd, we use 30 GHz RF amplifiers (SHF 807) and measure an electrical eye opening of 1.0 V, which operates the modulator in its linear regime. For the 77 GBd signal, we replace the 30 GHz RF amplifiers with 50 GHz amplifiers (SHF S807) and operate at an electrical eye opening of 1.4 V. The frequency response of the drive circuits and the RF cabling is determined in a reference measurement and digitally pre-compensated in the experiment. At the receiver, we use a 2 nm optical bandpass filter, an optical modulation analyzer (OMA, Keysight N4391) and 63 GHz real-time oscilloscopes (Keysight DSOZ634A), each having a sampling rate of 160 GSa/s.

**Experiment 2:** In Experiment 2, we use Micram DAC4 as a signal source. The device features an analog bandwidth of 40 GHz (or more) and a sampling rate of up to 100 GSa/s. The data pattern is derived from a pseudo-random pattern with length $2^{15}$. To generate 16QAM signals up to a symbol rate of 100 GBd, we use 70 GHz RF amplifiers (SHF 827). The drive voltage swing is 1.5 $V_{pp}$. For 80 GBd an 100 GBd, the signal is generated without oversampling. At the receiver side, we use a 1.5 nm-wide optical band pass filter, a coherent receiver (Tektronix OM4245), and 70 GHz real-time oscilloscopes (Tektronix DPO77002SX), featuring a sampling rate of 200 GSa/s.

## Acknowledgements

We acknowledge support and equipment loan from Micram, Keysight and Tektronix. We acknowledge A.K. Jen and J. Luo from Soluxra for providing SEO100 and SEO250. We further acknowledge support by the European Research Council (ERC Starting Grant 'EnTeraPIC', 280145), the EU-FP7 projects PhoxTroT (318240) and BigPipes (619591), the Alfried Krupp von Bohlen und Halbach Foundation, the Helmholtz International Research School for Teratronics (HIRST), the Karlsruhe School of Optics and Photonics (KSOP), and the Karlsruhe Nano-Micro Facility (KNMF).